\documentclass[conference]{IEEEtran}
\IEEEoverridecommandlockouts

\usepackage{booktabs}
\usepackage{hyperref}
\usepackage{multirow}
\usepackage{enumitem}
\usepackage[capitalise, noabbrev, nameinlink]{cleveref}
\usepackage[dvipsnames]{xcolor}
\usepackage{graphicx}
\usepackage{todonotes}
\usepackage{natbib}

\setcitestyle{numbers,square,comma} 
\def\BibTeX{{\rm B\kern-.05em{\sc i\kern-.025em b}\kern-.08em
    T\kern-.1667em\lower.7ex\hbox{E}\kern-.125emX}}
    \usetikzlibrary{calc}
    
\definecolor{labelColor}{RGB}{177, 74, 255}
\newcommand{\uilabel}[1]{\protect\tikz [font=\sffamily, baseline={($ (current bounding box.center) - (0,.3em) $)}] \fill[fill=labelColor] (0,0em) circle (0.6em) node[text=white] {#1};}

\begin{document}

\title{\textsc{Ender}: Visual Scaffolding for Geometric Proof Comprehension}

\author{
    \IEEEauthorblockN{
        Hwei-Shin Harriman\IEEEauthorrefmark{1}, 
        Yuchen Jin\IEEEauthorrefmark{2}, 
        Dominik Moritz\IEEEauthorrefmark{1}, and 
        Joshua Sunshine\IEEEauthorrefmark{1}
    }
    \IEEEauthorblockA{
        \IEEEauthorrefmark{1}Carnegie Mellon University---Pittsburgh, PA, USA
    }
    \IEEEauthorblockA{
        \IEEEauthorrefmark{2}University of Wisconsin---Madison, WI, USA
    }
}

\maketitle

\begin{abstract}
Geometric proof is a foundational but notoriously difficult topic in secondary mathematics, primarily due to the high cognitive load required to integrate visual diagrams with logical text. As proofs evolve, tracking the initial givens and an expanding set of derived truths can easily overwhelm working memory. While established scaffolding techniques are often visual and diagram-based, current digital learning environments mostly rely on static, text-heavy interfaces that force students to manage these complex mental models on their own. This paper introduces \textsc{Ender}, an interactive rendering system designed to visualize the logical status of two-column proofs. \textsc{Ender}’s design is grounded in four literature-derived requirements: (1) diagram annotation, (2) structural tracing, (3) support for partial or incorrect work, and (4) synchronization between the diagram and text. To evaluate the system, we conducted a between-subjects, think-aloud experiment with 12 secondary school students in the United States, comparing \textsc{Ender} against a traditional digital proof interface. Our results indicate that students using \textsc{Ender} performed significantly better on proof comprehension tasks—specifically in identifying and correcting logical errors. These findings suggest that dynamic visual scaffolding may ease the working memory burden of digital interfaces, offering a promising building block for future AI-based proof tutoring systems.
\end{abstract}


\section{Introduction}
Geometric proof is a key concept in secondary school mathematics, yet it remains one of the most difficult to learn~\cite{stylianou2009teaching, usiskin_van_1982, senk_how_1985}. A primary source of this difficulty is the high cognitive load required to coordinate two distinct representations: the visual diagram and the logical text. While the diagram provides spatial intuition, the text demands rigorous deduction. Novices frequently struggle to connect these modes, often viewing the diagram as a static picture rather than a dynamic representation of logical truths. They lose track of the evolving state of the proof, leading to errors where they confuse ``surface features''---such as the orientation or size of a shape---with the underlying logical structure~\cite{clements_geometry_1992, ulusoy_prospective_2021, fischbein_concepts_1998,Chazan_1993}.


To help students practice connecting diagram and text, teachers frequently rely on visual scaffolds, such as dynamic gestures and color-coded markings~\cite{dimmel2015semiotic,chen2013interplay}. Because proof diagrams are typically under-specified~\cite{laborde2005hidden}, these scaffolds serve as critical bookkeeping mechanisms that allow students to visually track logical information as it is established. This kind of dynamic marking comes naturally with pen and paper, where a diagram can be freely annotated as reasoning unfolds. While a major promise of digital proof tutoring systems is to support independent practice, most current environments lack this same functionality, instead restricting scaffolding and feedback to static text and symbols~\cite{tessier2017etude}. By stripping away these bookkeeping mechanisms, digital tools force students to manage this bookkeeping themselves, which makes it more difficult to focus on the actual logical deduction~\cite{purnama20, leppink2013development}.


To address this gap, we synthesized findings from learning sciences literature to derive four design requirements for an interactive rendering of proof: (1) annotate the diagram; (2) trace proof structure; (3) support partial and incorrect proofs, and (4) synchronize the diagram and textual proof. Based on these requirements, we developed \textsc{Ender}, an interactive rendering system that visualizes the logical status of a two-column proof. Through dynamic diagram annotations and color-coded highlighting, the system explicates dependencies between proof steps that are typically hidden in traditional formats, providing immediate feedback on full, partial, and incorrect proofs. \textsc{Ender} is architected not as a standalone tutor, but as a modular visual component. Its design can be embedded in diverse e-learning contexts---from problem-solving interfaces to analogical encoding tasks---serving as a complementary visual layer for AI-based feedback systems and intelligent tutors.

To evaluate our design, we conducted a between-subjects, think-aloud experiment to answer the research question: \textbf{Does adding \textsc{Ender} to two-column proofs result in improved accuracy on geometric proof comprehension tasks at the secondary school level?} Our experiment with 12 secondary school students in the United States compared performance on proof comprehension tasks using \textsc{Ender} against a traditional digital proof interface. Our results indicate that students using \textsc{Ender} were significantly better at proof comprehension tasks, particularly in identifying and correcting errors. This suggests that, by externalizing timely, relevant information, \textsc{Ender}'s visual scaffolding may ease the working memory burden that digital proof interfaces typically impose on students.

The rest of this paper is organized as follows: \cref{sec:background} establishes the relevant findings in learning science literature. We then describe design requirements motivated by that literature in \cref{sec:design-requirements}. We detail how \textsc{Ender}'s design meets each requirement in \cref{sec:system}. In \cref{sec:method} and \cref{sec:results} we describe the setup and findings of our between-subjects experiment. Lastly, we provide a summary of related tools and approaches in \cref{sec:related} and discuss the broader implications of \textsc{Ender} in \cref{sec:discussion}.

\section{Background}\label{sec:background}
This section describes the background literature that contributes to the design and motivation of \textsc{Ender}. Our discussion of related work, including existing tools and approaches, is deferred to \cref{sec:related}.
\subsection{Challenges caused by surface features}\label{sec:back-surface} 
Students consistently struggle to master the complexities of geometric proof~\cite{usiskin_van_1982,senk_how_1985}. This difficulty was quantified as early as 1985, when the National Council of Teachers of Mathematics reported that only 30\% of geometry students could produce acceptable solutions to proof-based problems~\cite{senk1985well}. Significant evidence suggests students' primary struggles with geometric proof are tied to their focus on surface features over deep structure~\cite{clements_geometry_1992, ulusoy_prospective_2021, fischbein_concepts_1998,Chazan_1993}. Most of these surface features are taken from the diagram included with the proof. While the diagram can help students visually track information, its role is deceptively complex. Certain low-level properties in a proof---such as between-ness or intersection---can be drawn directly from the diagram, whereas higher-level properties---such as congruence or perpendicularity---can only be concluded using a deductive argument~\cite{herbst2013geometric,laborde2005hidden}. This often murky distinction tempts students to place blind faith in the diagram. Many students incorrectly assume it is drawn to scale and derive conclusions based on relationships that visually appear to be true~\cite{Soucy_McCrone_Martin_2004a,cirillo2019addressing}. They mistake the drawing for the concept, so when a drawing's surface features change, their reasoning breaks down even if the underlying logic is identical. For example, students who learn about triangle congruence with both triangles in the same orientation may struggle when the triangles are flipped or overlapping. By mistaking the orientation---a surface feature---for a defining property, they fail to recognize the shapes are congruent and which theorems can be applied~\cite{weber2001student}. This limited perspective prevents students from grasping the full generality of proof. 

\subsection{Proof is a non-linear, deductive process}\label{sec:back-nonlinear}
Geometric proof also marks an abrupt cognitive shift from computational to formal deductive reasoning. Prior curricula in arithmetic and algebra emphasize procedural problem-solving, where success is defined by a correct answer. Geometry demands a new paradigm: students are no longer asked what the answer is, but why a statement must be true. This transition from empirical observation (“it looks true”) to formal argument---where conclusions are necessary consequences of established logic---is a common struggle~\cite{stylianou2009teaching}. Many students, believing measurement is sufficient proof~\cite{ahmadpour_students_2023,Chazan_1993,selden_overcoming_2008,mccrone2004assessing}, find this first formal encounter with abstract deduction uniquely challenging and distinct from their mastered computational skills. 

Many proofs can be approached in multiple ways. While experts are comfortable with pursuing multiple lines of reasoning, students are far more inclined to attempt to solve proof sequentially~\cite{karunakaran2018need}. This new way of thinking is daunting even for more advanced students: undergraduates often struggle to identify an appropriate starting point for a proof~\cite{moore1994making} or fail to understand which definitions and theorems they can invoke as they work~\cite{weber2001student}. The most common format, the two-column proof, also contributes to these misunderstandings. Its linear structure simplifies the underlying graph-shaped logical argument of a proof~\cite{ball_teaching_2003}. Introducing students to alternate formats, such as flowchart~\cite{miyazaki2014functions} or paragraph, can help them visualize the true structure. \citet{anwar2021effect} found that students trained on multiple formats demonstrated significantly better logical reasoning and understanding of the critical ideas of proof. However, most of the proofs that students encounter are---at least in the US---given in the two-column format, including testing and textbook materials~\cite{herbst2002establishing}.

\subsection{Use of the diagram in geometric proof}\label{sec:back-diagrams}
Despite the risk of being misled by surface features, the diagram plays an essential role in geometric proof. High-achieving geometry students habitually mark their diagrams with known information, valid assumptions, and refer to the diagram to plan their proofs~\cite{cirillo2021competencies}. \citet{koedinger1990theoretical} observed that experts rely on the diagram for proof. They seek out high-level patterns or ``schemas'' that help them to quickly sort out what is known and what can be concluded. This lets them find paths through the proof in fewer steps than a novice. Our formative interviews with geometry instructors also found that teachers emphasize a similar pattern-matching behavior with their students, teaching them to mark their diagrams and pick out visual patterns to progress. 

\subsection{Challenges of tracking proof state}\label{sec:back-tracking}
Beyond conceptual hurdles, students face incidental challenges in managing the proof’s dynamic state. A proof is an evolving process where each new statement, once justified, becomes a new premise for subsequent steps. Students must maintain a high cognitive load, tracking the initial ``given'' information and an expanding set of derived truths. For example, after proving triangles congruent, they must update their mental model to include that all corresponding parts are now congruent. This demanding bookkeeping, which is extraneous to logical deduction, can overwhelm working memory, obscure the argument's logical structure, and hinder planning~\cite{purnama20, leppink2013development}. Even high-achieving students fail to grasp the evolving logical status of proof. They tend to complete proof ritualistically by constructing arguments that receive the highest marks without understanding why the logic demands that specific structure~\cite{healy2000study}.

\subsection{Challenges of providing robust scaffolding and feedback}\label{sec:back-feedback}

Many of the challenges that students face in proof can be addressed with high-quality feedback that directs attention, supports metacognitive regulation, and connects outcomes to underlying principles~\cite{kluger1996effects,narciss2013designing}. In the classroom, teachers are expected to provide personal feedback to students and correct their errors quickly. However, this relies upon a number of factors~\cite{lesseig2016investigating}, including the teacher's understanding of what constitutes a proof~\cite{knuth2002secondary} and the types of activities they assign~\cite{martin2005interplay}. This challenge is exacerbated by the lack of resources available to teachers, whose textbooks frequently emphasize memorization and low-level cognitive tasks~\cite{sears_examination_nodate}. Despite the best efforts of teachers, the number of proof-related misconceptions a student could harbor are practically limitless. Even students in the same class can demonstrate different levels of understanding~\cite{ramandaniinvestigating}, making it difficult to attend to each student's current skill simultaneously.

Producing high-quality automatic feedback comes with its own set of challenges. Because the validity of a proof depends on its logical structure, it cannot be checked with the same techniques as arithmetic or algebra problems. \citet{tessier2017etude} surveyed \emph{Intelligent Tutoring Systems} (ITS) specifically designed for geometric proof. These systems predominantly rely on pre-coded sets of solutions which are either defined by experts (for which there are multiple options per proof)~\cite{anderson1985geometry,agentgeom,koedinger1990abstract} or derived using automatic deduction~\cite{matsuda2004gramy,leduc2016qed-tutrix,baghera} to provide corrective feedback. Any additional explanatory or personalized feedback is presented as text or audio that is predefined by an expert~\cite{tessier2017etude}. Only a few systems are capable of providing feedback related to the student's interaction with a diagram, such as ANGLE~\cite{koedinger1990abstract}, \textsc{Agent-Geom}~\cite{agentgeom}, or \textsc{Cabri-DEFI}~\cite{cabri-defi}. The majority only offer feedback based on the student's logical argument and cannot address any misconceptions related to the visual component of the proof unless the expert predicted them.

\begin{figure*}[ht!]
\centering
\includegraphics[width=.9\textwidth]{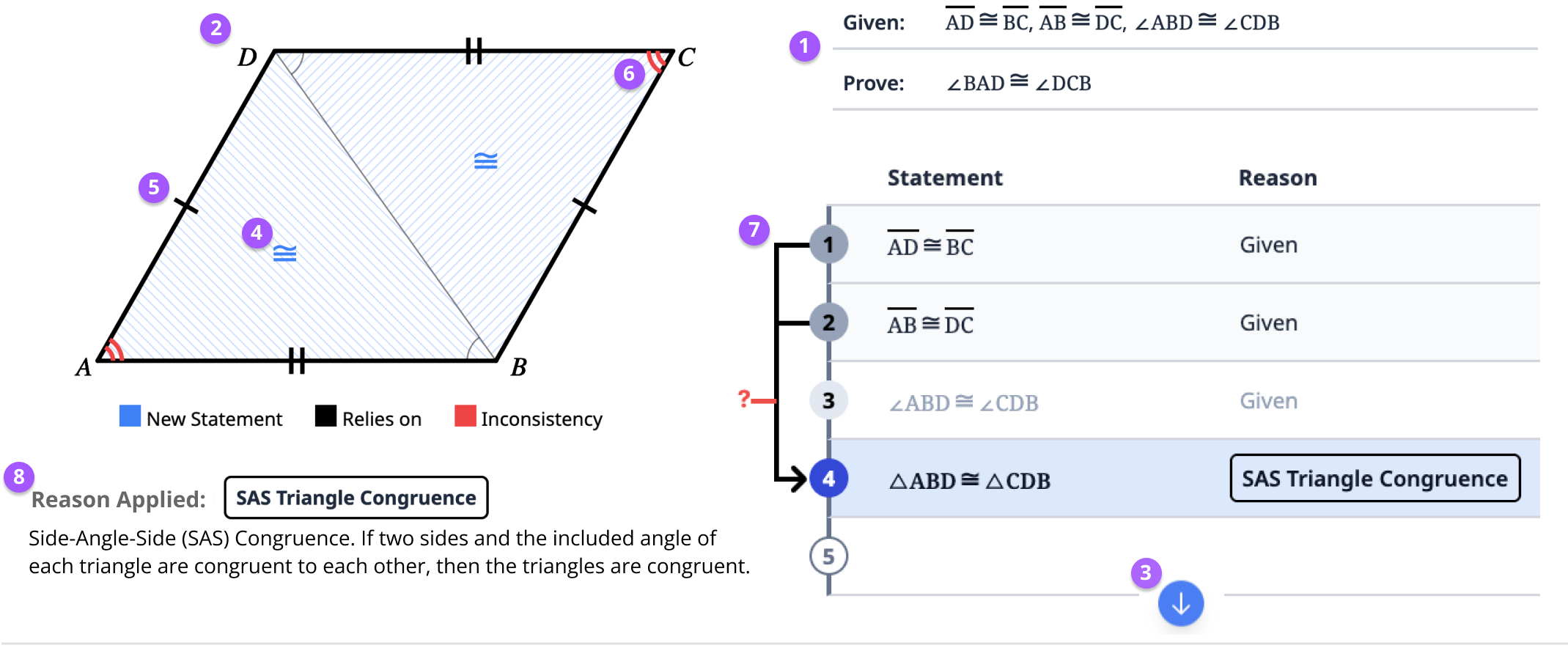}
\caption{\textit{Top to bottom}: The \textsc{Ender} interface displaying the 4th step of the example proof from \cref{fig:example}. Students navigate by clicking or using the Up/Down arrow keys. \uilabel{1} Premises for the proof. \uilabel{2} Interactive construction, updates at every step. \uilabel{3} Indicator that proof has not been fully revealed. \uilabel{4} Blue highlights the new statement. \uilabel{5} Black highlights prerequisites for the current reason. \uilabel{6} Red highlights inconsistencies between the current step's prerequisites and proof state. \uilabel{7} Points to prerequisites' location(s) in proof, ``?'' means one is missing. \uilabel{8} Text definition of current reason.} 
\label{fig:ender-interface}
\end{figure*}

\section{Design Requirements for \textsc{Ender}} \label{sec:design-requirements}
In this section, we present our motivating design requirements for the development of \textsc{Ender}. We defer discussion of technical implementation to \cref{sec:system}.

We had to make many design decisions during the development of \textsc{Ender}. Due to the high sustained cognitive load required to complete proof, the interface should highlight key information without overwhelming users. To guide our process, we developed a set of considerations to inform when and how information is presented to the student. These considerations were informed by the challenges described in \cref{sec:background}, then refined through our experiences across multiple design iterations and pilots. For each design requirement, we refer back to the relevant research from \cref{sec:background}.

\textbf{DR1: Annotate diagram.} As discussed in \cref{sec:back-surface}, diagrams provided with geometric proof are typically underspecified. While low-level properties can be drawn from the diagram, high-level properties, like congruence, must be deduced~\cite{laborde2005hidden}. Marking the diagram with known information helps students to clarify these ambiguities~\cite{cirillo2021competencies}. Automatically annotating the diagram to match the current proof state ensures that visual information is presented only after it has been logically justified. This can direct the student’s attention and more clearly tie the current state of the diagram to the proof. 

\textbf{DR2: Trace proof structure.} This requirement is motivated by the challenges introduced by the nonlinearity of proof (\cref{sec:back-nonlinear}). Students often struggle to understand why a proof must be constructed in a certain order~\cite{healy2000study}. There is some evidence that representing proofs as a flowchart improves student reasoning~\cite{miyazaki2014functions,anwar2021effect}. \textsc{Ender} draws from the design of flowchart proofs to expose the connections between proof steps that are normally hidden by the standard two-column format.

\textbf{DR3: Support partial and incorrect proofs.} This requirement is derived from the need to provide high-quality feedback on proofs at scale (\cref{sec:back-feedback}). Feedback is most helpful when provided as the student encounters issues, such as on incomplete and incorrect proofs. To provide feedback in these situations, the underlying proof checker must recover from errors wherever possible and continue providing feedback on other parts of the proof~\cite{keenan2024learner}. Any accompanying rendering should also visualize errors in the proof's argument. Even if there is a mistake early in the proof, \textsc{Ender} follows these guidelines and tries to provide feedback on the rest of the proof as though the incorrect step was valid.

\textbf{DR4: Synchronize diagram and textual proof.} As noted in \cref{sec:back-tracking}, proof is an evolving process where students must track an expanding set of derived truths. This is a cognitively demanding management task. Filtering the information students see to match exactly what is necessary at each step in the proof could help scaffold this process. To fulfill this requirement, \textsc{Ender} allows the student to navigate to any step in the proof to view relevant information about its state and structure. 



\begin{figure}[t!]
\centering
\includegraphics[width=.75\columnwidth]{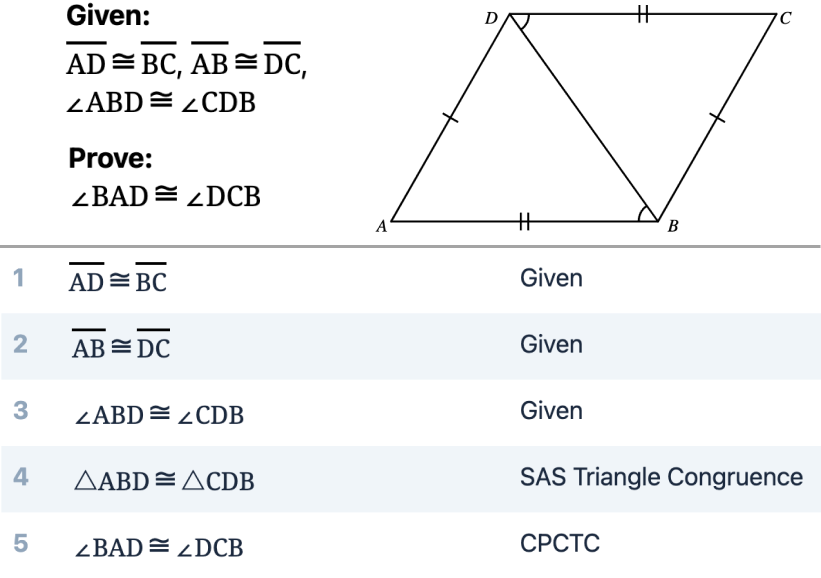}
\caption{An example two-column geometric proof. The proof is incorrect because it uses SAS Triangle Congruence without enough prerequisites. Adding the statement $\overline{BD} \cong \overline{BD}$ by the Reflexive Property before step 4 would correct this proof.} 
\label{fig:example}
\end{figure}

\section{System Design}\label{sec:system}

Based on our design requirements, we developed \textsc{Ender}, an interactive rendering technique for geometric proofs. \textsc{Ender} presents a synchronized, side-by-side view of a textual proof and its corresponding geometric construction, highlighting key information as students progress through the logical sequence. Designed as a modular component, it is intended to be integrable into diverse digital e-learning environments.

\subsection{Augmenting the Two-Column Rendering}\label{sec:two-column}

While educational researchers have explored various alternative proof formats \cite{miyazaki2014functions, ye09}, the two-column proof remains the undisputed pedagogical standard in secondary mathematics curricula \cite{Winer2022Investigating, otten14}. Consequently, \textsc{Ender} is designed to augment, rather than replace, this traditional rendering. Understanding \textsc{Ender}’s design thus requires a clear definition of the standard two-column format, an example of which is shown in \cref{fig:example}.

Every two-column proof includes a few key pieces of information: the premises and the goal to prove. These are typically accompanied by a construction: a diagram illustrating the geometric scenario. The proof itself consists of sequential steps comprising statements and reasons. In a mathematically sound proof, each statement must be logically deducible from its assigned reason in combination with previously established statements.

In the example provided in \cref{fig:example}, the student attempts to prove triangle congruence, but the proof is logically flawed. Specifically, the ``Side-Angle-Side Triangle Congruence'' (SAS) reason cited in Step 4 is inapplicable; SAS requires an \textit{included} pair of congruent angles. As written, the first three statements correspond to a Side-Side-Angle (SSA) configuration, which is insufficient to guarantee congruence. To rectify this proof, a student would need to use the Reflexive Property to establish that $\overline{BD} \cong \overline{BD}$ prior to Step 4.

\subsection{Fulfilling Design Requirements}

\textsc{Ender} augments the standard rendering to meet our specific pedagogical goals. \Cref{fig:ender-interface} displays a partial rendering of the example proof within the \textsc{Ender} interface.

\subsubsection{Annotating the Construction}\label{sec:system-annotating}
To bridge the gap between abstract statements and geometric intuition, \textsc{Ender} provides a dynamic interactive construction (\uilabel{2}). When a user selects a step in the proof, the system automatically renders the relevant geometric properties directly onto the diagram. For example, if a statement asserts that two segments are congruent, the construction is updated with standard geometric markings (e.g., hash marks). This visual feedback ensures that students can immediately perceive the geometric consequences of each logical claim.

\textsc{Ender}'s annotations follow the standard rules that students learn: objects with the same type (i.e., a pair of segments) and number of ticks are tied to each other. \textsc{Ender} supports standard segment and angle congruence marks, right angle marks, segment similarity marks, and parallel segment marks. In case multiple distinct sets of objects are congruent, \textsc{Ender} automatically renders the appropriate number of tick marks for each one. 

In secondary-level geometry, there is no standard way to mark two triangles as congruent. During pilots, students were thrown off when statements involving segments or angles were annotated, but steps classifying triangles showed no equivalent visual change. To solve this in \textsc{Ender}, a ``$\cong$'' symbol is rendered in the center of each triangle.

The annotations relevant to the selected step are highlighted with the following color scheme: \textbf{\textcolor{MidnightBlue}{blue}} for annotations corresponding to the new statement, \textbf{black} for the reason's prerequisites, and \textbf{\textcolor{BrickRed}{red}} for any inconsistencies. Inconsistencies are determined by comparing the current proof state to the definition of the applied reason. 

In \Cref{fig:ender-interface} blue congruence symbols and triangle shading \uilabel{4} correspond to the current statement $\triangle ABD \cong \triangle CDB$. The black segments \uilabel{5} corresponding to step 4's dependencies ($\overline{AD} \cong \overline{BC}$ and $\overline{AB} \cong \overline{DC}$). Finally, two angles, $\angle DAB$ and $\angle DCB$ are highlighted in \textbf{\textcolor{BrickRed}{red}} \uilabel{6} because those angles are expected dependencies of the SAS Triangle Congruence theorem used in Step 4. This visualization helps students verify whether the necessary conditions---such as two sides and an included angle---have actually been established in the diagram. 

\subsubsection{Tracing Proof Structure}\label{sec:system-tracing}

\begin{figure*}[ht!]
\centering
\includegraphics[width=.85\textwidth]{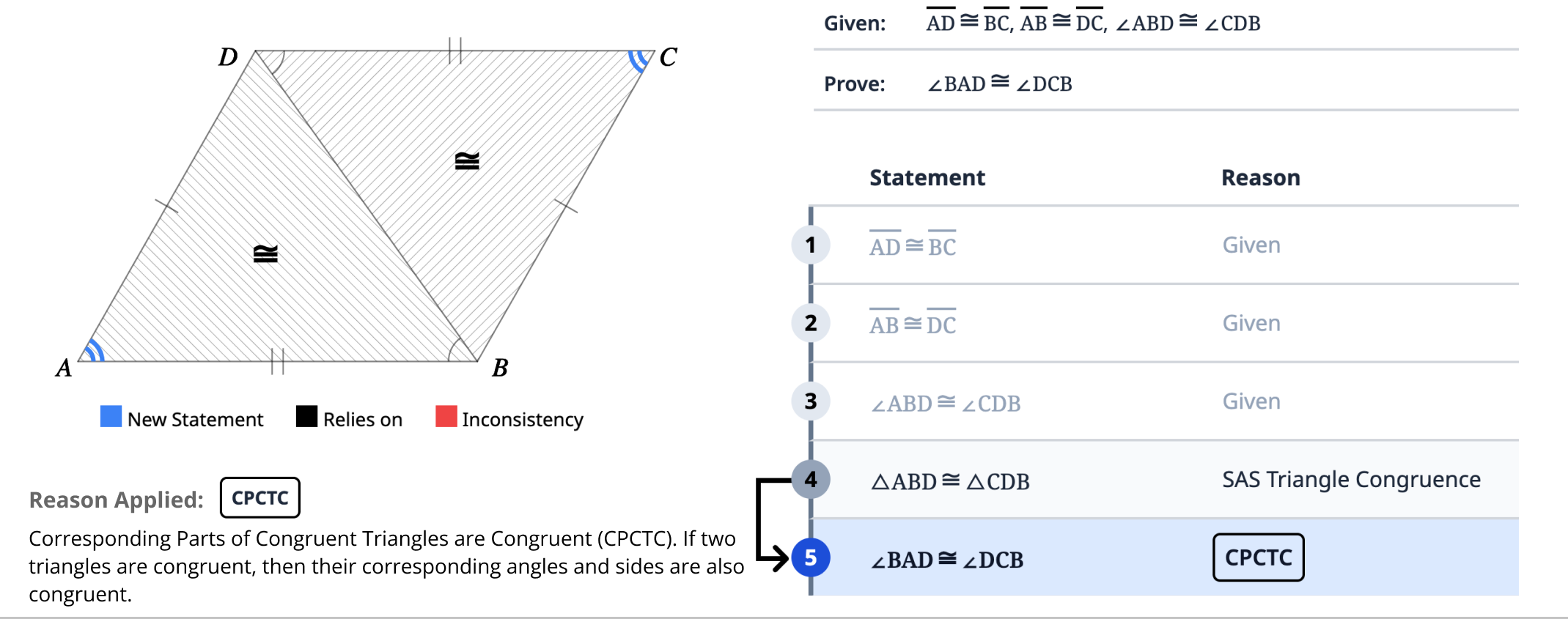}
\caption{The \textsc{Ender} interface for the 5th step of our example proof.} 
\label{fig:step5}
\end{figure*}

In \cref{fig:ender-interface}, statements of steps 1 and 2 are  dependencies for step 4. \textsc{Ender} renders dependencies via arrows between steps placed to the left of the textual proof. The arrows are designed to emphasize the logical connections between steps in a proof and follow the same logic as a flowchart proof~\cite{miyazaki2014functions}. The number of arrows matches the requirements of the theorem; for instance, SAS requires arrows between three steps, while in step 5 (\cref{fig:step5}) ``Corresponding Parts of Congruent Triangles are Congruent'' (CPCTC) only requires one.

If dependencies are missing from the proof (e.g., trying to use SAS but missing the angle pair), a broken node with a red \textcolor{BrickRed}{``?''} is rendered. There is no meaning behind the vertical placement of the question mark relative to other proof steps. When there are multiple incomplete ways to resolve a dependency, \textsc{Ender} will highlight issues with the earliest and most complete set. For instance, both steps 1, 2 and steps 2, 3, are equally likely to contribute to SAS. Since steps 1 and 2 appear earlier in the proof the dependencies are drawn there. Dependencies can only be drawn to earlier steps in a proof. If a step breaks this rule (e.g., if CPCTC was used before SAS), it is rendered as a missing dependency. 

This dependency information can be used by students to ``debug'' a proof. For example, if a student is unsure why  $\angle DAB$ and $\angle DCB$ are red, they can check the \textbf{``relies on''} arrows connecting steps 1 and 2 to step 4 \uilabel{7}. These indicate that the first two steps are prerequisites for step 4. The presence of a \textbf{\textcolor{BrickRed}{red ``?''}} signals that some information is missing. By reviewing the highlights, previous steps, and definition \uilabel{8}, the student is able to confirm that SAS Triangle Congruence cannot be used because the red angles have never been declared congruent.

\subsubsection{Supporting Partial and Incorrect Proofs}
\textsc{Ender} evaluates the validity of each step in isolation, allowing students to proceed even if previous steps contain errors. If a student encounters an incorrect claim, the system visualizes the error by rendering the conflicting tick marks on the diagram, rather than preventing the action. This persistence is intentional: it treats the proof as a ``debugging'' task. By seeing the red, conflicting state of the diagram, the student is cued to resolve the logical inconsistency with geometric reasoning. However, the visual conflict signals an issue without guaranteeing a solution. For example, in \Cref{fig:ender-interface}, the system highlights the visual conflict caused by a missing step. The most direct solution is to introduce the reflexive property ($\overline{BD} \cong \overline{BD}$), which is \textbf{not} highlighted, requiring the student to deduce the missing property independently.


\subsubsection{Synchronizing Diagram and Textual Proof}
The core innovation of \textsc{Ender} is the interactive rendering that keeps the construction and the text in a synchronized loop. As the student navigates through the steps using the interface or keyboard, the ``state'' of the proof updates globally. The current statement being examined is highlighted in blue in the text while simultaneously being emphasized in the construction. Furthermore, a text definition of the current reason is provided alongside the visual highlights, ensuring the student is presented with a unified representation of what the reason requires, what the proof has established, and how those facts appear on the geometric construction.

\Cref{fig:step5} shows the \textsc{Ender} interface for step 5 of the proof. If you compare to \Cref{fig:ender-interface}, you will notice that many elements of the interface have changed. Step 5 is highlighted in blue in the textual proof, indicating that it is the current step. There is an arrow from Step 4 to Step 5 since Step 4 is the only dependency of CPCTC. The diagram has also changed. $\angle BAD$ and $\angle DCB$ are highlighted in blue because those angles now correspond to Step 5's statement. Similarly, the congruence markers are highlighted in black because triangle congruence is the (consistent) dependency of the CPCTC reason applied in Step 5. All of the visual bookkeeping tasks usually required of students are applied by \textsc{Ender} automatically to scaffold their understanding of the proof. 

\subsection{Implementation}
\textsc{Ender} is implemented as a web application developed using React and TypeScript. It can support proofs involving the common object types in secondary-level Euclidean plane geometry, including points, lines, segments, angles, triangles, circles, and polygons. It has not been tested on other types of geometry, such as deductive database methods~\cite{chou2000deductive}, or three-dimensional geometry. It also does not currently support proofs that involve algebraic manipulation, such as, proving that one segment is a fraction of another. To date, we have implemented 12 correct and incorrect proofs using 13 different statements and 20 reasons. 

Since the front-end rendering of proof is the focus of this study, the current implementation has no true understanding of the proof's state. Instead, it takes a detailed specification of the proof's evolution and outputs the interactive rendering as a webpage. The specification includes basic information about the proof's setup as well as step-specific information. Basic proof information includes the coordinates and names of each point in the diagram, the givens and goal, and a list of proof step objects. Each step includes the type of statement and reason used, references to any other steps it relies on, and information about the color, visibility, and type of tick mark for every object in the diagram. This information is used to generate the diagram and dependency arrows with styling that is updated depending on the currently focused step in the proof.

\begin{figure*}[ht!]
\centering
\includegraphics[width=.85\textwidth]{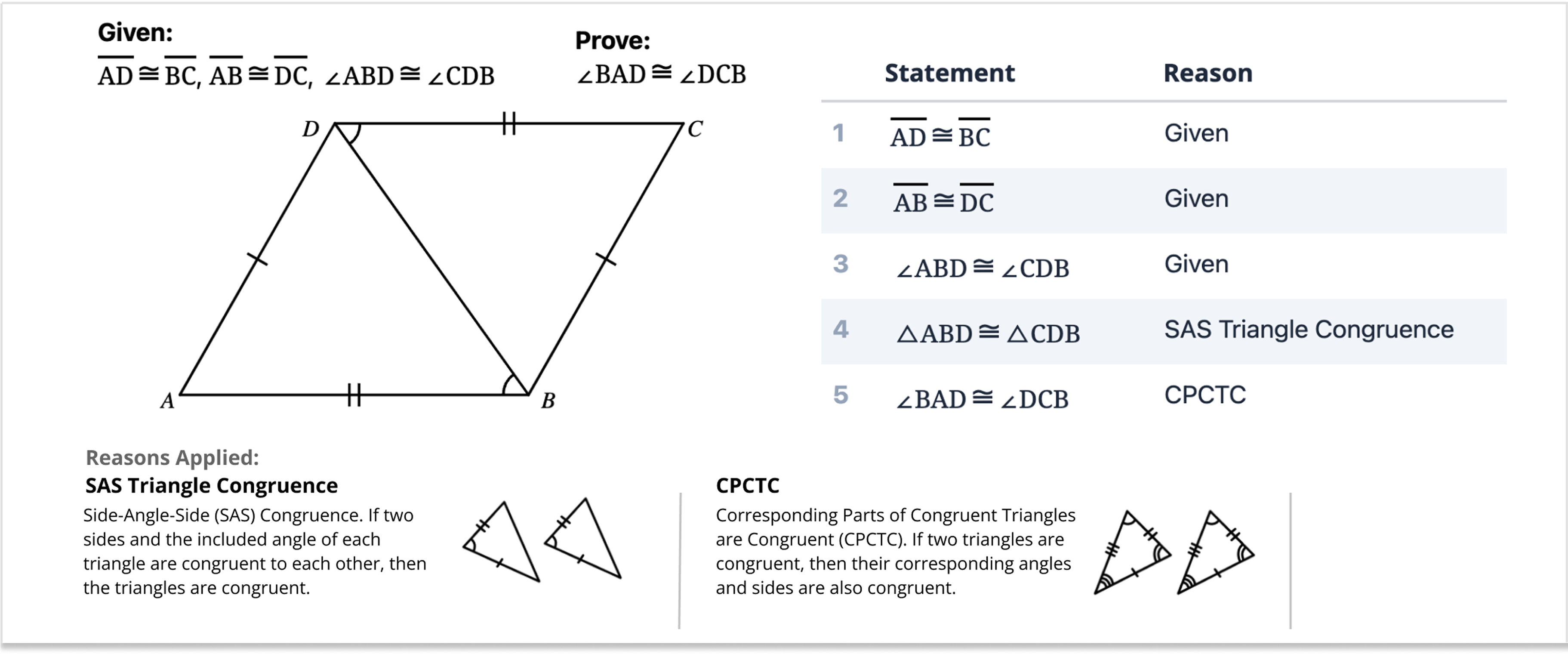}
\caption{Screenshot of the interface used by participants in the control condition. It provides all the standard information included in a two-column proof as well as illustrated definitions for each reason applied.} 
\label{fig:interfaces}
\end{figure*}

\section{Method} \label{sec:method}

To evaluate the effectiveness of \textsc{Ender}, we conducted a between-subjects think-aloud experiment comparing it against a traditional proof format. Our primary dependent measure is task accuracy. We hypothesize that by externalizing the logical state of a proof, \textsc{Ender} will enable students to achieve higher accuracy than those using a standard interface. The following sections detail our experimental conditions, procedure, and evaluative tasks.

\subsection{Conditions}\label{sec:conditions}
We compared the \textsc{Ender} interface (\cref{sec:system}) against a control interface (\cref{fig:interfaces}). The control condition provides a digital rendering of a traditional two-column proof that is stylistically similar to \textsc{Ender} but maintains the standard presentation found in paper-based curricula. In this condition, tick marks on the diagram only reflect the ``given" information, and illustrated definitions for geometric reasons are displayed separately from the proof steps. This design mirrors the typical student experience of cross-referencing a textbook, where definitions are often isolated from the problem context.

The primary functional differences in the control are the absence of dynamic highlighting, interactivity, and explicit dependency arrows. Furthermore, while \textsc{Ender} diagrams dynamically accumulate tick marks for each justified statement—including any incorrect assertions—the control diagram remains static. We considered providing the control group with a diagram matching the final state, but pilot results showed this misled students to assume all statements were logically valid.

\subsection{Tasks}
For each task, participants first reviewed a proof to familiarize themselves with its structure and logic. They then answered two to three questions selected from a bank of 38 predefined items designed to assess proof comprehension. These items were informed by established assessments for geometric proof comprehension~\cite{Koedinger_1991,yang2008model,Soucy_McCrone_Martin_2004a} and refined through multiple pilot rounds. The questions required students to evaluate existing proofs by considering their generality, correctness, or logical status. Each question in the bank was categorized into one of five targeted types:  \begin{enumerate}[label=\textbf{QT\arabic*},topsep=.1em]
\item \textbf{Error Identification and Correction}:
\begin{itemize}
\item\label{qt:1} \textbf{A}: Is there a mistake in this proof? \item \textbf{B}: What is the first step that is wrong? \item \textbf{C}: If the proof can be corrected without changing the givens, describe the necessary changes (2 mins max).
\end{itemize}
\item\label{qt:2} \textbf{Logical Dependency} : Is there enough information to apply {reason $R$} at step {$S1$} between steps {$S2$} and {$S3$}? 
\item\label{qt:3} \textbf{Conclusion Validation} : By the end of the proof, has enough information been established to correctly conclude that {object $O1$} is congruent to {object $O2$}? 
\end{enumerate}  

\subsection{Procedure} \label{sec:procedure}
This study received approval from the Institutional Review Board (IRB) of Carnegie Mellon University. Since our participants were under the age of 18, written informed consent was obtained from a legal guardian, and each participant gave written assent before participation. All sessions were audio and screen recorded. Experiments were conducted with a mouse, keyboard, and 1440p monitor. Participants interacted with a React-based web application that presented proofs and captured responses via a top-mounted toolbar.
 
Each student participant was randomly assigned to one of the two conditions. Participants were instructed to complete the tasks to the best of their ability while thinking aloud~\cite{chi1994eliciting}, and encouraged to interact freely with the interface, provide educated guesses, and use provided scratch paper. To capture a real-time indicator of current focus, participants were required to keep their hand on the mouse throughout the procedure~\cite{huang2012user}. 

The experimental session was structured into four distinct phases: \textbf{1) Pretest (10 min):} Established baseline proficiency in fundamental congruence concepts, standard notation, and triangle theorems (e.g., SAS, SSS).
\textbf{2) Orientation (5 min):} A researcher-led tutorial introduced the interface features and the think-aloud protocol using two example proofs. This phase served strictly for familiarization and was excluded from the final performance analysis.
\textbf{3) Think-Aloud Tasks (35 min):} Participants completed eight geometric proofs using their assigned interface. A 7-minute limit was enforced per proof, and no accuracy feedback was provided during the session.
\textbf{4) Exit Interview (10 min):} A semi-structured interview where participants reflected on their experience and any encountered difficulties.

Participants viewed ten proofs: two for the tutorial (one correct, one incorrect) and eight for the main procedure (three correct, five incorrect). Diagrams featured intentional symmetry to identify students relying on perceptual cues rather than logical deduction~\cite{clements_geometry_1992, ulusoy_prospective_2021, fischbein_concepts_1998}. This design probed whether participants would base assertions on a figure’s appearance---such as assuming a triangle is equilateral---rather than its established logical status~\cite{Soucy_McCrone_Martin_2004a}. For example, the triangles in \cref{fig:interfaces} appear equilateral, inviting students to conclude side or angle equivalence without formal justification. All participants evaluated the eight experimental proofs in a randomized sequence.

Students initially reviewed each proof to induce a priming effect~\cite{strack1992order}, encouraging engagement during the think-aloud process. Each evaluation began with an error-identification question (\ref{qt:1}\textbf{-A}). If a mistake was identified, participants were prompted to pinpoint (\ref{qt:1}\textbf{-B}) and resolve it (\ref{qt:1}\textbf{-C}). Participants then answered one to two additional questions regarding alternative proof sequences (\ref{qt:2}) or the final state of the proof (\ref{qt:3}). Depending on the errors reported, participants answered between 22 and 38 questions in total. This think-aloud approach was designed to reveal student reasoning patterns and pinpoint specific misconceptions~\cite{chi1994eliciting}.

\begin{table}
\caption{Participant demographics for each condition. Age, Grade, and Years Since Geometry (YSG) were recorded at the time of participation. A ``0'' in YSG indicates that the student was currently taking geometry.}
\label{tab:participants}
\begin{tabular}{@{}l|lrrr@{}}
                   & \textbf{Participant} & \textbf{Age} & \textbf{Grade} & \textbf{Years Since Geometry}  \\* 
\midrule
\multirow{7}{*}{\rotatebox[origin=c]{90}{\textsc{Ender}}} & SP1          & 15  & 10    & 0                     \\*
                   & SP3          & 15  & 10    & 1                     \\*
                   & SP5          & 15  & 10    & 1                     \\*
                   & SP7          & 15  & 10    & 0                     \\*
                   & SP9         & 16  & 11    & 2                     \\*
                   & SP11         & 14  & 9     & 0                     \\* 
\midrule
\multirow{6}{*}{\rotatebox[origin=c]{90}{Control}} & CP2          & 16  & 10    & 1                     \\*
                   & CP4          & 17  & 12    & 3                     \\*
                   & CP6          & 15  & 10    & 1                     \\*
                   & CP8          & 17  & 12    & 3                     \\*
                   & CP10          & 15  & 10   & 0                     \\*
                   & CP12         & 15  & 9     & 0                    
\end{tabular}
\end{table}

\subsection{Participants} Twelve U.S. secondary school students were recruited through school networks and referrals for in-person sessions conducted from November 2024 to February 2025 in Pennsylvania and California. Participants were required to be U.S.-based and either currently enrolled in or recently graduated (within two years) from a geometry course with triangle congruence proficiency. All participants were familiar with the two-column format. \cref{tab:participants} summarizes participant demographics. Balanced prerequisite knowledge was confirmed by mean pretest scores (81\%; Ender SD=11\%, Control SD=7\%). Participants received \$20 gift cards upon completion.


\subsection{Scoring}
Scoring was uniform across conditions, with each correct answer earning one point. The three-part \ref{qt:1} question was weighted at $1/3$ point per sub-part; specifically, \ref{qt:1}\textbf{-C} required a valid correction without extraneous steps or the correct identification of an uncorrectable proof. This balanced weighting prevents error-detection tasks from disproportionately skewing the accuracy results. Total accuracy was calculated as the ratio of points earned to the total available points.

\section{Results} \label{sec:results}

\begin{figure}
\centering
\includegraphics[width=\columnwidth]{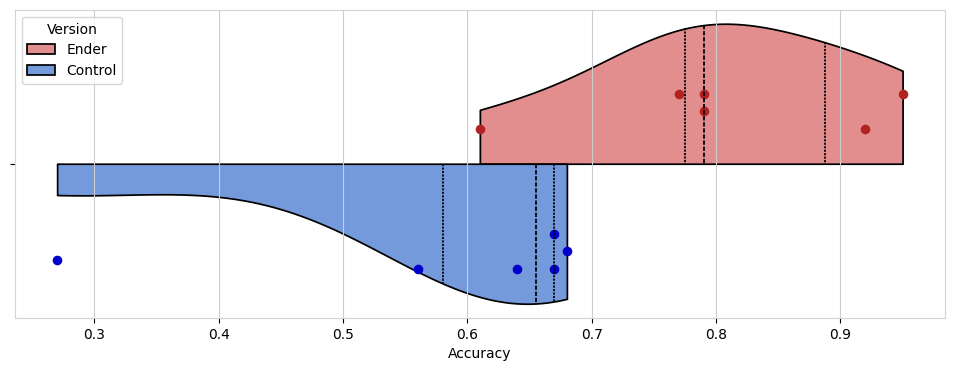}
\caption{Accuracy distributions by interface. \textsc{Ender} users achieved significantly higher mean accuracy than the control group (80\% vs. 58\%).} 
\label{fig:scoreViolin}
\end{figure}

In this section, we describe the results of our between-subjects experiment. First, we present the quantitative results, analyzing the scores of each group to answer our research question. Then, we present qualitative results of the think-aloud transcripts to demonstrate the impact of \textsc{Ender} on verbalized geometric reasoning.

\textbf{Quantitative Results:} Our research question examines the effect of \textsc{Ender} on proof comprehension. We used a linear regression model (ANCOVA) to determine the effect of version (\textsc{Ender} vs control interface) and pretest scores on overall accuracy. Version had a significant effect on overall accuracy (Estimate=$.223$, $p<0.03$). Pretest scores were not significant (Estimate $=0.01$, $p=.97$). Students referring to \textsc{Ender} had a mean accuracy of 80.5\% (SD=.12), while those referring to the control only achieved a mean accuracy of 58.2\% (SD=.16). The distribution of accuracies for each interface is visualized in \Cref{fig:scoreViolin}.

\begin{figure}
\centering
\includegraphics[width=.9\columnwidth]{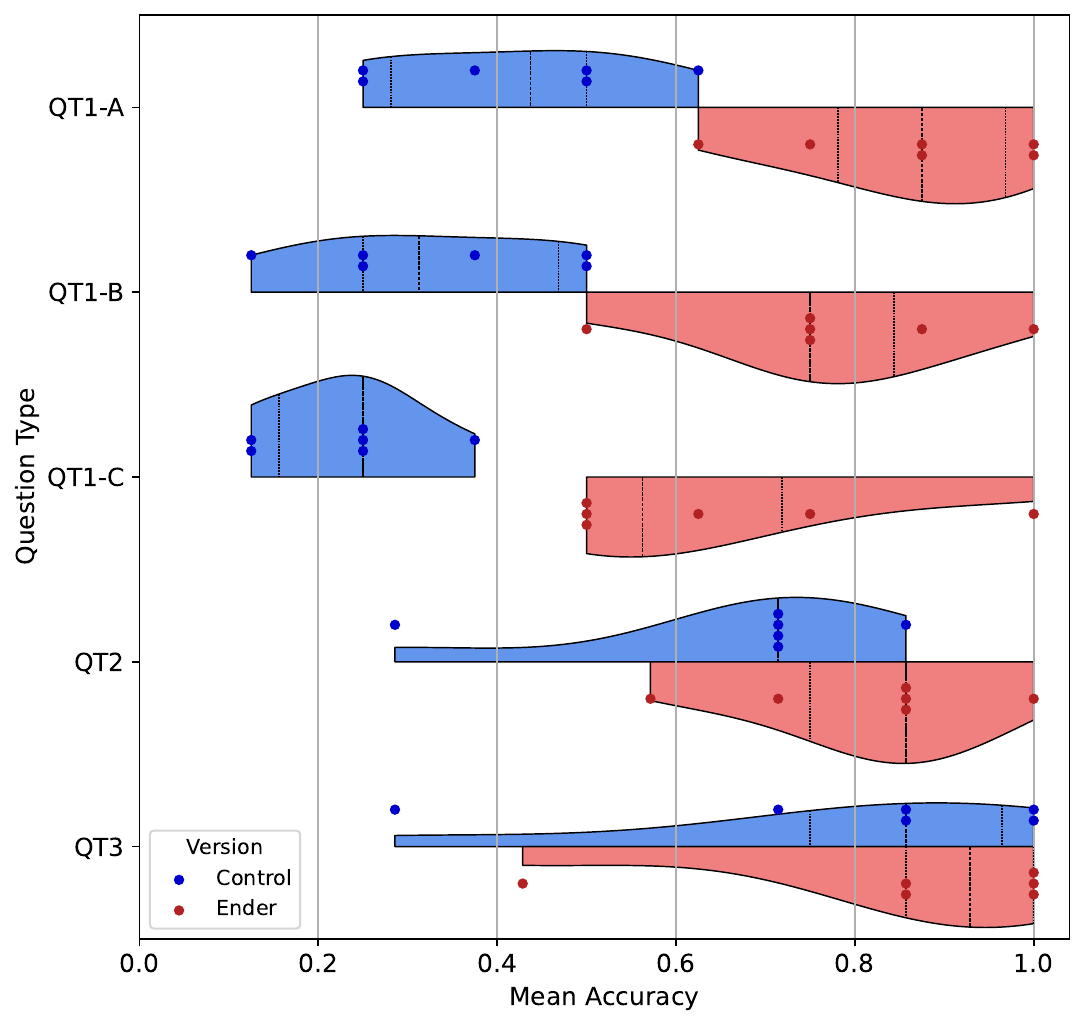}
\caption{Raincloud plots visualizing the distribution and per-participant accuracy for each of the 5 question types. Participants in the \textsc{Ender} group were significantly more accurate on questions about finding and fixing mistakes (all of \ref{qt:1}), and slightly more accurate on questions about logical dependencies (\ref{qt:2}).} 
\label{fig:questionTypeViolin}
\end{figure}

We also analyzed performance across the five question types. We used a linear mixed-effects model to analyze the effects of version (\textsc{Ender} vs control interface), question type, and their interaction on participant scores. We controlled for pretest scores as a covariate and accounted for random intercepts by participant. Again, pretest scores were not a significant predictor of final performance ($F(1,9)=0.014$, $p = 0.91$). We found significant main effects of version ($F(1, 9) = 16.64$, $p <0.005$) and question type ($F(4, 40) = 12.40$, $p < 0.0001$), indicating that performance varied across question types. A significant interaction between version and question type was observed ($F(4, 40) = 4.36$, $p<0.005$), indicating that the performance advantage of \textsc{Ender} relative to the control interface varied by question type. Specifically, participants using \textsc{Ender} performed significantly better on \textbf{QT1-A} (Estimate = $0.36, p<0.005$), \textbf{QT1-B} (Estimate = $0.36, p<0.005$), and \textbf{QT1-C} (Estimate = $0.30, p<0.01$). These question types involve identifying and correcting mistakes. The raincloud plot in \Cref{fig:questionTypeViolin} shows the differences in mean accuracy for each of the five question types.

\textbf{Analysis of Geometric Reasoning:} To determine whether students answered based on geometric reasoning, we first created a rubric of acceptable justification for every problem. Next, the first author qualitatively coded all of the think-aloud transcripts and screen recording data. When a student answered a question, we noted whether they provided a verbal justification using geometric terms to describe why a statement is true or false. If the student voiced multiple justifications, their final verbalization was used for coding. For instance, with the proof in \Cref{fig:example}, an example of valid geometric reasoning for the question, ``Is there a mistake in this proof?'' would be, ``Yes, because there is not enough information to use SAS Triangle Congruence at step 4.'' 1 point was given for each demonstration of correct reasoning. If students provided what could be a valid justification in a special case (i.e., in \Cref{fig:example}, claiming that ``all of the angles are equal'' is valid if and only if the triangles are known to be equilateral), but the special case was not established within the scope of the proof, then it was marked as incorrect reasoning. In the \textsc{Ender} group, students who verbally acknowledged a visual cue were given 1 point only if they correctly explained the cue's meaning in the context of the proof using geometric terms (e.g., in \Cref{fig:ender-interface}, pointing to the \textbf{\textcolor{BrickRed}{?}} and explaining that there isn't enough information to use SAS). Answers submitted without any verbalization received 0 points. 

Our analysis revealed that students in both conditions answered with some form of geometric reasoning in 90\% of tasks. The other 10\% (factored into the correlation analysis as incorrect reasoning) included unknown reasoning or random guesses. We used a Pearson correlation analysis~\cite{cohen2013applied} to evaluate the relationship between each student's accuracy and their demonstrated reasoning. The results revealed a strong positive correlation between the two variables, $r = 0.954$, $t(10) = 10.113$, $p < 1.4 \times 10^{-6}$, indicating a significant relationship between geometric reasoning and accuracy. This correlation shows that the correctness of a student's answer was closely tied to their verbalized reasoning. Since students using \textsc{Ender} achieved significantly higher overall accuracy than control group users, this means they also demonstrated a higher rate of correct geometric reasoning.

\section{Related Work}\label{sec:related}
Research on digital tools in geometry education has largely focused on \emph{Dynamic Geometry Environments} (DGE) such as \textsc{GeoGebra}~\cite{geogebra}, \textsc{Geometer's Sketchpad}~\cite{gsp_5}, or \textsc{Desmos}~\cite{desmos}. These environments allow students to construct and manipulate figures, a process known to significantly improve their ability to mentally model geometric properties~\cite{sinclair2012technology, weigand2025geometry,laborde2002integration}. However, these tools are not designed to support the student through formal deductive reasoning. \citet{komatsu2020interplay} found that students, when provided with both a DGE and pen-and-paper, use DGEs to discover properties through construction but switch to pen-and-paper to draft and work through the actual proof. 


This disconnect may stem from a lack of visual support for the logical layer of mathematics. \citet{head_math_2022} analyzed the common types of mathematical augmentations that content creators use to communicate complex ideas, identifying techniques such as variable line weights, color-coding, and explicit linking between text and figural properties. ~\citet{butcher2013using} found that students are better at correcting their mistakes in geometry problems when feedback is placed in proximity to relevant components of the diagram. Unfortunately, these supportive features are rarely found in geometry education tools. Tools like IXL~\cite{ixl_learning} and \textsc{Student Desmos}~\cite{student_desmos} offer scribble tools on some of their activities, enabling students to freely mark over a diagram. However, they lack the semantic awareness to enforce or provide feedback on the correctness of those marks. On the other hand, platforms like \textsc{DeltaMath}~\cite{korzyk_deltamath_2026}, \textsc{Géométrix}~\cite{gressier_geometrix}, and \textsc{Advanced Geometry Tutor}~\cite{matsuda2005advanced} can highlight specific geometric objects referenced in a step (e.g., highlighting segment $AB$ when the student types ``AB''). These features are often implemented as simple locators rather than as tools for logical scaffolding. 

The closest antecedent to our work is \textsc{ANGLE}~\cite{koedinger1990theoretical}, which intentionally uses visual cues to guide students toward valid geometric patterns. However, \textsc{ANGLE} teaches students to use experts' high-level patterns that do not align with Euclidean rules found in textbooks. This requires specialized instruction that does not easily transfer to standard curricula. In contrast, \textsc{Ender} systematically renders the valid, invalid, and dependent states of the proof using Euclidean logic. This approach allows the system to support the standard curriculum while still providing the visual scaffolding that \citet{head_math_2022} advocates.

Researchers have also explored altering the proof format itself to improve comprehension, investigating flowcharts~\cite{koedinger1990abstract, anderson1985geometry,chypre1992}, paragraph proofs~\cite{mentoniezh}, and multi-stage or hybrid approaches~\cite{cabri1997,leduc2016qed-tutrix}. Despite this exploration, the two-column format remains the dominant standard in US classrooms, and there is still no conclusive evidence that implementing alternative formats in digital tools yields significantly better learning outcomes. Consequently, rather than enforcing a novel format, \textsc{Ender} is designed to augment the existing two-column standard, overlaying the dependency structure visually to reveal the logic typically hidden by the text.

\begin{figure*}[ht!]
\centering
\includegraphics[width=.85\textwidth]{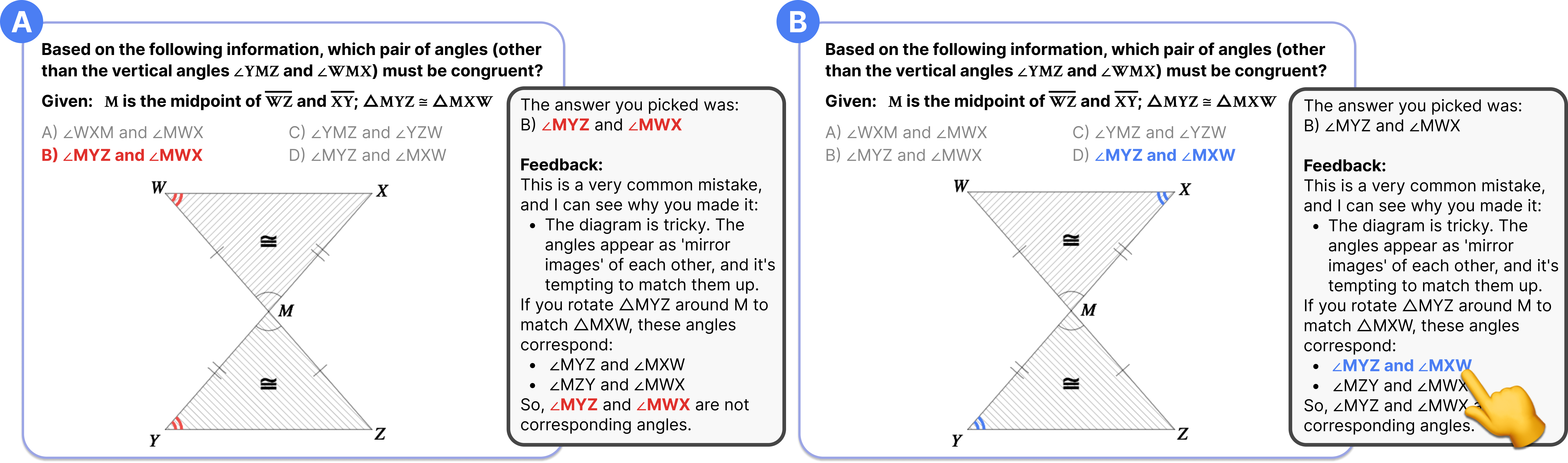}
\caption{\textsc{Ender} feedback integration mockups. Selective color visually links textual feedback to the construction: (A) red highlights identify errors in both text and diagram; (B) blue highlights appear on hover to signify correct elements.} 
\label{fig:feedback}
\end{figure*}

Finally, our approach parallels intelligent tutors in other domains that expose the underlying state of complex systems. In programming education, \textsc{Python Tutor}~\cite{guo2013online} renders the runtime stack and heap alongside source code, while the \textsc{Whyline} system~\cite{ko2004designing} links program output directly to the logic that produced it. Similarly, in physics, the \textsc{Andes} tutor~\cite{vanlehn2005andes} uses color-coded vectors to flag incorrect force components, helping students debug their intuition. \textsc{Ender}'s rendering can be thought of as an application of this ``runtime visualization'' strategy to geometric proof: by rendering the logical state of the proof interactively, it allows students to debug their geometric reasoning much as a programmer debugs code.

Ultimately, we view the visualization of logical state not as a competing pedagogical strategy, but as a fundamental interface requirement for digital proof. In \cref{sec:discussion}, we elaborate on how \textsc{Ender} can complement the current ecosystem of geometry tutors, offloading the cognitive burden of tracking the proof so that students can focus on the logic itself.

\section{Limitations}

Our experiment has several limitations. First, because students used \textsc{Ender} as a single interface rather than variants with individual features removed or altered, we cannot attribute the improved accuracy and reasoning to any specific design element. Second, our design did not capture high-resolution data on student process: although we collected mouse-movement data as a proxy for visual attention, it was too coarse to fully model students' cognitive processes with and without \textsc{Ender}. Third, because we measured performance in a single session, we cannot tell whether gains reflect durable comprehension or a short-term effect consistent with the assistance dilemma~\cite{koedinger2007exploring}, where scaffolding aids performance without necessarily producing lasting learning. Fourth, our proof comprehension test, though piloted extensively, has not been validated by other geometry experts. Finally, our statistically significant results come from a small sample, so we caution against generalizing broadly from this study. Future work should examine \textsc{Ender}'s cognitive effects directly and run larger, ablation-based studies to isolate which features matter most.

\section{Discussion and Future Work}\label{sec:discussion}

We introduced \textsc{Ender} to investigate whether an interactive, visually scaffolded rendering could support student comprehension of geometric proofs. Our experimental results indicate that \textsc{Ender} significantly enhances proof comprehension accuracy, particularly in tasks requiring the identification and correction of errors. Furthermore, the strong positive correlation between task accuracy and correct geometric reasoning suggests that \textsc{Ender}'s visual cues are externalizing bookkeeping information which in turn, helps students focus on the logical structure of the proof. Beyond these primary findings, our results and design have broader implications for the design of intelligent tutoring systems, classroom instruction, and alleviating the cognitive burden of proof comprehension.

\subsection{Reducing Cognitive Load in Proof Comprehension}
In addition to our primary analysis, we noticed some interesting qualitative trends that warrant further investigation. In both groups, the most frequent error type involved making inferences based on an incorrect recollection of which geometric properties had already been established. Students made this mistake 42 times (60\% of all errors in the control group), compared to only 13 times (35\% of all errors) in the \textsc{Ender} group. This suggests that the standard two-column interface without any augmentation is underspecified for the task at hand. In fact, we observed that control students frequently used mouse movements and hand gestures to simulate the missing visual information. Furthermore, despite being provided with scratch paper, most students attempted to solve problems solely within the interface. Only four students, three in the control and one in \textsc{Ender}, used paper. Two of the control students incorrectly reproduced their figures in a new orientation, while the others sketched unlabeled subsections of a diagram. This reluctance to switch mediums suggests that the cost of transferring work to and from the digital workspace is too high. By integrating state-tracking directly into the diagram, \textsc{Ender} seems to offload some of this cognitive burden, allowing students to focus their limited resources on logical deduction rather than memory management. These findings align with previous critiques that the static two-column format provides insufficient support for novice learners~\cite{Soucy_McCrone_Martin_2004a}.

\subsection{Applying \textsc{Ender} to Other Proof Formats}
While we have evaluated \textsc{Ender} as an augment to the two-column format, we believe that its core design components could be fitted to the other two common formats: flowchart and paragraph. The problem statement and definitions of each reason are format-independent, leaving three major components to \textsc{Ender}: the diagram, the drawn dependencies between steps, and the step-by-step navigation through the proof. The diagram component, including its color-coding and highlighting, extends naturally to both formats. The dependency arrows are built into flowchart proofs, so incorrect incoming connections to a step could simply be highlighted. Paragraph proofs are more difficult because their open-ended format relies heavily on specific phrasing and word choice. \textsc{Ender} could present the dependencies by drawing connections between the current statement and previous sentences, but would likely need specific UI adjustments (inline highlights, sidebars, colored text) to make the connections visually obvious. To navigate a flowchart proof, \textsc{Ender} can pick any order that respects the dependencies (e.g., traversing left-to-right, top-to-bottom). Paragraph proofs, on the other hand, have no inherent step boundaries, so \textsc{Ender} would need to segment the text by phrase or sentence to enable navigation.

\subsection{Complementing E-Learning Platforms}
While \textsc{Ender} was evaluated as a standalone comprehension tool, its primary value proposition lies in its potential to complement existing e-learning platforms, including ITS. Current geometry ITS feedback is predominantly text-based~\cite{tessier2017etude}, requiring students to read an explanation and mentally map it onto a complex diagram. We believe \textsc{Ender}'s rendering engine can complement both hard-coded and generative feedback using Large Language Models (LLMs) by serving as a visual connector for textual feedback. As illustrated in \Cref{fig:feedback}, \textsc{Ender} could be integrated into feedback loops to selectively highlight the geometric components referenced in an error message. For instance, if a student incorrectly cites a theorem, an ITS could use \textsc{Ender} to visually contrast the student's claim (highlighted in red) with the actual geometric state (highlighted in blue). This approach aligns with the principle of contiguity in multimedia learning~\cite{mayer2017using}, reducing the split-attention effect~\cite{ayres2005split} by placing the visual cue directly on the object of study.

\subsection{Implications for Instruction and Scaffolding}
Beyond automated tutoring, future work will expand \textsc{Ender}’s capabilities to test new forms of instructional scaffolding. Specifically, we are developing a grammar and proof checker to automatically render secondary-level proofs in the \textsc{Ender} style, moving away from hard-coded states for faster, more scalable task creation. For example, experts could easily create partial proof problems where specific visual cues are toggled on or off to isolate different skills. A faded scaffolding approach~\cite{mcneill2006supporting} could initially provide full visual support---highlighting both the statement and its logical dependencies---and then gradually remove these cues as the student demonstrates mastery. \textsc{Ender} could also be used in classroom instruction. Instead of a teacher manually drawing and erasing markings on a whiteboard---a process that is ephemeral and hard for students to track---\textsc{Ender} allows for a structured, reversible step-through. This ensures that the visual representation of the argument remains consistent and accessible to all students, regardless of the pace of the lecture.

\section{Acknowledgments}
The authors would like to thank Paulo Carvalho and Peter Cheng for their advice regarding the statistical modeling and experimental design. This material is based upon work supported by the National Science Foundation under Award Nos. 2447499, 2346174, and 2119007, as well as Grant No. DGE2140739. Any opinions, findings, and conclusions or recommendations expressed in this material are those of the author(s) and do not necessarily reflect the views of the National Science Foundation. Funding was also received from the ARCS (Achievement Rewards for College Scientists) Foundation, Pittsburgh Chapter.

\bibliography{refs}

\end{document}